# Jets, Blazars and the EBL in the GLAST-EXIST Era

Jonathan E. Grindlay and the *EXIST* Team

*Harvard-Smithsonian Center for Astrophysics, 60 Garden St., Cambridge, MA 02138*

**Abstract.** The synergy of GLAST and the proposed *EXIST* mission as the Black Hole Finder Probe in the Beyond Einstein Program is remarkable. With its full-sky per orbit hard X-ray imaging (3-600 keV) and "nuFnu" sensitivity comparable to GLAST, *EXIST* could measure variability and spectra of Blazars in the hard X-ray synchrotron component simultaneous with GLAST (~10-100GeV) measures of the inverse Compton component, thereby uniquely constraining intrinsic source spectra and allowing measured high energy spectral breaks to measure the cosmic diffuse extra-galactic background light (EBL) by determining the intervening diffuse IR photon field required to yield the observed break from photon-photon absorption. Such studies also constrain the physics of jets (and parameters and indeed the validity of SSC models) and the origin of the >100 MeV gamma-ray diffuse background likely arising from Blazars and jet-dominated sources. An overview of the *EXIST* mission, which could fly in the GLAST era, is given together with a synopsis of other key synergies of GLAST-*EXIST* science.

## INTRODUCTION

Scaling from the results of the EGRET telescope on the Compton GRO mission, the dominant extragalactic source population that will be detected and studied in detail with GLAST are almost certainly the Blazars. These are the luminous compact sources in the nuclei of active galaxies with jet-dominated emission in which relativistic beaming plays a dominant role in defining the apparent luminosity and variability of a source region (or regions) in which non-thermal particles are accelerated (presumably in shocks) and produce synchrotron radiation with broad spectral distribution extending from the radio to hard X-ray bands. The gamma-ray spectrum (defined here as >1 MeV) is likely produced by inverse Compton emission of the synchrotron electrons on their own photons or on external photon fields (e.g. from the host galaxy or spatially distinct emission regions such as an accretion disk at the base of the jet. For Blazars, with strong polarization and power law spectra dominated by synchrotron emission, the "synchrotron self-Compton" (SSC) models (e.g. Grindlay 1975, Band and Grindlay 1986, Krawczynski, Coppi and Aharonian 2002) have achieved considerable success in explaining spectral shapes and variability of Blazars as well as less luminous jet-dominated active galactic nuclei (AGN) systems. However, the spectacular variability of Blazars, especially at TeV (but also ~10-100GeV, accessible to GLAST) energies, makes it clear that high sensitivity and high throughput correlated observations at X-ray/hard X-ray energies are needed to both constrain SSC models and test whether competing hadronic models are appropriate in some cases.

The wide-field, broad band and imaging Energetic X-ray Imaging Survey Telescope (*EXIST*) mission (Grindlay 2005), under study as the Black Hole Finder Probe (BHFP) in NASA's Beyond Einstein Program, is the perfect complement to GLAST to provide unique constraints on the physics of Blazars, other classes of extreme non-thermal sources containing relativistic jets (and, in particular, cosmic Gamma-ray Bursts, GRBs, one of the key science objectives of *EXIST*), as well as a host of other objectives. Here we first introduce the *EXIST* mission concept, as recently defined for the NRC Beyond Einstein Program Assessment Committee (BEPAC), and then summarize how a period of joint operations with GLAST would not only probe the physics of Blazar jets but allow their broad-band and time-variable spectra to be used to measure the still poorly-known flux and spectrum of the extragalactic background light (EBL) as manifest in the diffuse cosmic IR background. This, in turn, allows unique constraints on the integrated (over cosmic time) star light and thus nuclear luminosity of the Universe, complementing another key *EXIST* science objective to constrain the accretion luminosity of the Universe by measuring the obscured AGN fraction in an all sky survey. We also outline several other fundamental science objectives that would benefit from simultaneous GLAST-*EXIST* operations.

## OVERVIEW OF *EXIST*

An overview figure showing the *EXIST* mission concept is shown in Fig. 1 along with a Table of key mission parameters. The mission is a scanning coded aperture imager, with two large area telescope arrays (HET and LET) covering the nearly identical fully-coded fields of view shown. Each of the 19 sub-telescopes of the HET is built up from a close-tiled array (56 x 56cm) of imaging Cd-Zn-Te (CZT) detectors (each 2 x 2 x 0.5cm, with 1.25mm pixels in a 16 x 16 array read out by a pair of 128ch. ASICs) that view coded masks at 1.4m focal length for 6.8arcmin resolution and 56arcsec source positions (90% confidence radius, 5σ survey threshold sources). The LET, with finer pixels (160µ), provides finer positions (11arcsec, 90% confidence radius) for unique AGN identifications with what may appear as "normal" galaxies at the survey flux threshold of ~0.05mCrab (=5 x $10^{-13}$ erg/cm$^2$-sec in band dE=E/2), or comparable to the all-sky sensitivity of the ROSAT survey at 0.3-2.5 keV. With this sensitivity, logN-logS studies (from Chandra) of AGN and the likely ~70% obscured fractions (e.g. Treister and Urry 2005) suggest a total AGN survey sample of >3 x $10^4$ of which at least ~500 will be Blazars.

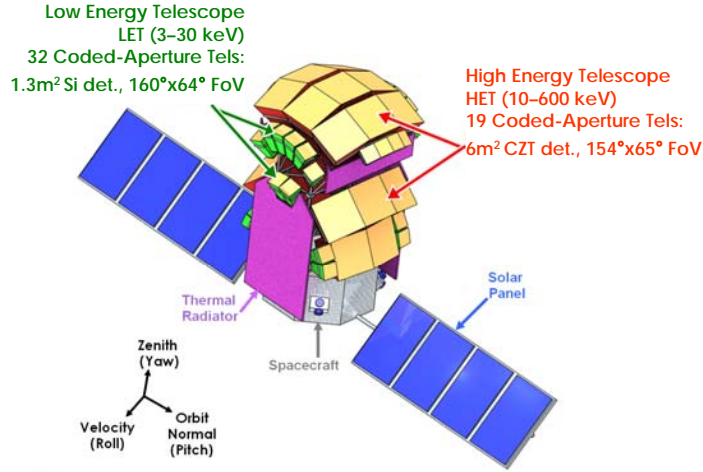

| Survey | No. Obj. |
|---|---|
| AGN | >3. E4 |
| GRBs | 700/y |
| **Flux lim** | **Full sky** |
| dE=E/2 in: | 1y (cgs) |
| 3-100 keV | 5 E-13 |
| ≤600 keV | 5 E-12 |
| **Time Res.** | **Units** |
| ea. Photon | 0.1msec |
| GRB pos. | 10sec |
| Mission dur. | 5y |
| **En. Resol.** | **dE/E (%)** |
| 3-600 keV | 1 – 5% |

**FIGURE 1.** Layout of *EXIST* showing the two large area coded aperture telescopes, zenith pointed on a scanning spacecraft.

## *EXIST*-GLAST SYNERGY FOR BLAZARS, JETS AND THE EBL

Because *EXIST* is zenith pointed with its very large FoV, and in addition "nods" ±12° each ~15min towards the orbit poles, the *full sky is uniformly covered each 95min orbit*, or at twice the rate covered by GLAST. Thus the two missions are highly complementary in their sky coverage and sensitivity for Blazars, with unpredictable flares and changing spectral parameters. This is of course equally true for other highly variable non-thermal sources, usually involving emission from jets, such as GRBs in particular. We focus here on the Blazars for which the broad-band X-ray to Hard X-ray/soft gamma-ray coverage of *EXIST* provides the ideal complement spectral/temporal coverage for both GLAST and ground-based TeV (e.g. VERITAS) coverage.

The SSC model is the preferred model for lower-luminosity "blue Blazars" which have their low-energy peaks (in their ν $F_ν$ spectra) in the X-ray range and their high-energy peaks in the Gev-TeV energy range. This in contrast to the case of the higher-luminosity "Red Blazars," with low-energy peaks in the EUV range and high-energy peaks at MeV-GeV energies, where upscattering of photons external to the jet is usually invoked to explain the large gamma-ray fluxes (e.g., Sikora, Begleman, & Rees 1994). The spectral dichotomy between "blue vs. red" Blazars is shown in Fig. 2 with approximate sensitivity limits for *EXIST* vs. GLAST and VERITAS. Although the simple division of Blazars into these two classes (Fossati et al 1998) now seems to be a selection effect (e.g., see Giommi et al. 2005), it is still the case that the "low-luminosity" Blazars and corresponding BL Lac type objects are those most likely described by SSC models: their underlying accretion is most likely a radiatively inefficient ADAF (or associated variant) which represents a relatively unimportant source of seed photons for inverse Compton scattering.

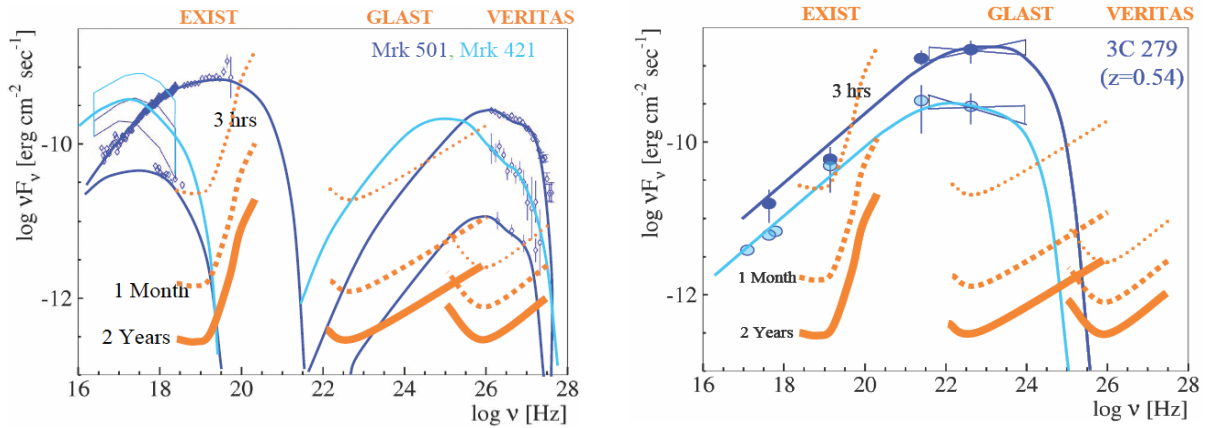

**FIGURE 2:** "Blue" (left; both synchrotron and IC components shown) vs. "Red" (right; only IC shown) Blazars showing the relative spectral sensitivities and constraints that can be provided by *EXIST* vs. GLAST and VERITAS (from Krawczynski 2004). The *EXIST* sensitivity now (with the LET) is relatively constant down to ~$10^{18}$ Hz.

The blue, SSC Blazars are of most interest for constraining the diffuse cosmic IR background with *EXIST* because a *single* population of high-energy electrons is then responsible for the low energy synchrotron emission observed by *EXIST* and the high-energy Compton-upscattered photons observed at gamma-ray energies (*EXIST* observations can thus constrain the intrinsic gamma-ray spectra of blue Blazars). Furthermore, these Blazars emit gamma-rays with energies up to ~20 TeV, allowing us to use $\gamma+\gamma \rightarrow e^+ e^-$ processes to probe the extragalactic background light (EBL) in the broad wavelength range from 0.5 to 25 microns. The broad-band 3-600 keV spectral coverage of *EXIST* constrains the synchrotron spectral break for blue Blazars and thus allows the IC spectrum measured by GLAST to fix the *intrinsic* spectral break of the IC spectrum. This is the essential measurement for the spectral break actually measured by VERITAS to be used to constrain the EBL. By carrying out this time-dependent measurement on many Blazars at a range of redshifts and lines of sight, the EBL can finally be unfolded with sufficiently redundant checks to constrain the diffuse cosmic IR background and thus the nuclear luminosity of the Universe. The same studies will provide powerful tests of the underlying SSC vs. external Compton models and will break the degeneracies between leptonic and hadronic models by broad spectral coverage in the time domain.

Similarly complementary spectral/temporal imaging of GRBs with *EXIST* vs. GLAST would constrain the jet physics of these remarkable events: the spectral sensitivity of *EXIST* for GRBs extends up to ~10MeV by using the large area active shields and would provide high statistics time-resolved spectral coverage to complement the GLAST-GBM as well as provide <10arcsec real-time GRB positions from the HET and LET imaging telescopes. The remarkable synergy of GLAST with *EXIST* for studies of extreme physics could be realized if *EXIST* were to be launched early in the Beyond Einstein mission queue.

## ACKNOWLEDGMENTS

I thank Henric Krawcznski and Paolo Coppi for their help and work on Blazar science for *EXIST*. This work was supported in part by NASA grant NNG04GK33G.## REFERENCES